\documentclass[11pt]{article}
\usepackage{graphicx}
\begin{document}

\title{Oscillation of Pauli Paramagnetism in Rotating Two-Component Fermionic Atom Gases}

\author{Beibing Huang and Shaolong Wan\thanks{Corresponding author.
Electronic address: slwan@ustc.edu.cn} \\
Institute for Theoretical Physics and Department of Modern Physics \\
University of Science and Technology of China, Hefei, 230026, {\bf
P. R. China}}

\maketitle
\begin{center}
\begin{minipage}{120mm}
\vskip 0.8in
\begin{center}{\bf Abstract} \end{center}

{By rotating two-component fermionic atom gases in uniform
magnetic field, a similar physical situation with de Haas-van
Alphen effect is constructed. We calculate magnetic moment of the
system and find that owing to an existence of effective magnetic
field coming from the rotation, the magnetic moment also shows the
oscillatory behavior about magnetic field, but it is completely
different from the famous oscillation of de Haas-van Alphen
effect. This distinction is due to that in the atomic gases the
orbital motion of atom only couples to rotation and does not
contribute to magnetic moment in the light of atomic charge
neutrality.}

\end{minipage}
\end{center}

\vskip 1cm

\textbf{PACS} number(s): 03.75.Ss, 05.30.Fk, 32.10.Dk
\\
\\
\\
Ultracold fermionic quantum gases are versatile and robust systems
for probing fundamental condensed-matter physics problems owing to
their highly controllability and operability \cite{greiner,
kinoshita, demarco, regal}. What happens to an s-wave BCS
superfluid when the numbers of up and down spin become unequal?
This is a 40-year-old problem which remains unresolved today.
However by controlling the particle number of different spin
species, experiments \cite{zwierlein, partridge} in atomic Fermi
gases are beginning to yield information and have provided access
to the exotic superfluid phases such as breached-paired phase
\cite{Sarma, Liu}, phase separation \cite{P., Caldas} and
Flude-Farrel-Larkin-Ovchinikov phase \cite{lo, ff}. By overlapping
some pairs of counter-propagating laser beam from different angle,
a hexagonal lattice is built \cite{zhao} so that the relativistic
Dirac fermions can be simulated and observed. Moreover, by tuning
the lattice anisotropy, one can realize both massive and massless
Dirac fermions and observe the phase transition between them
\cite{shi}. Besides these rotating superfluid fermion system is
also attracting. The effect of the rotation on the behavior of a
superfluid is also a longstanding subject of investigation in
condensed-matter physics \cite{rj}. Because of the irrotationality
constraint imposed by the existence of the order parameter a
superfluid cannot rotate like a normal fluid. Only when the
external angular velocity exceeds a certain critical value, it
will be energetically favorable to permit a quantum vortex to
enter into the system \cite{bruun}, which provides the ultimate
proof that the system has undergone a transition to a superfluid
state \cite{pitaevskii, hemmer}.

In condensed-matter physics in the presence of strong magnetic
field and low temperature, the formation of discrete Landau level
becomes significant in metals and is responsible for many
phenomena that some physical quantities show an oscillatory
behavior as a function of inverse magnetic field such as de
Haas-van Alphen effect \cite{callaway, abrikosov}. For an electron
system, the appearance of magnetic field acts on the degrees of
freedom of spin and orbital motions, but in atomic gases only
magnetic moment of atom experiences the magnetic field which leads
to the absence of Landau level. The rotation not only affects the
superfluid, but also leads to an effective magnetic field for
atomic gases as shown in the literature \cite{oktel, cooper}. So
the introduction of both magnetic field and rotation at the same
time will make cold atomic system be equivalent to an electron
system in magnetic field. In fact, this equivalence is very
superficial and the magnetic oscillation shows a completely
different scenario as shown below.

In this brief paper, we calculate the magnetic moment in rotating
fermionic atom gases. The orbital motion of neutral atoms is not
affected by the external magnetic field and only atomic magnetic
moment experiences it. So only Pauli paramagnetism exist. We find
that the magnetic moment shows an completely different oscillatory
dependence on the external magnetic field from the de Haas-van
Alphen effect.

We consider an idealized two-component fermionic atom system in an
external magnetic field $B$. In the frame of reference rotating
with angular velocity $\omega\hat{z}$, the Hamiltonian for a
particle of mass $m$ in an harmonic trap of natural frequency
$\omega_0$ only in $xy$ plane is$\colon$
\begin{eqnarray}
H=\frac{\vec{p}^{\,2}}{2 m} + \frac{1}{2}m\omega_0^2(x^2+y^2) -
\frac{g}{2}\mu_B B \sigma_z-\omega\hat{z}\cdot \vec{r}\times
\vec{p} \label{1.1}
\end{eqnarray}
with $\mu_B=\frac{e\hbar}{2m}$, $g$ and $\sigma_z$ being Bohr
magneton, atomic Lande factor and Pauli matrix respectively. The
rotation term can be rewritten so that the structure of Landau
level appears
\begin{eqnarray}
H=\frac{1}{2m}(\vec{p}+e\vec{A})^2-\frac{g}{2}\mu_B B
\sigma_z+\frac{1}{2}m(\omega_0^2-\omega^2)(x^2+y^2) \label{1.2}
\end{eqnarray}
with$A_x=m \omega y/e$ and $A_y=-m \omega x/e$. This form
indicates the effect of rotation is partitioned into two different
parts. The last term in (\ref{1.2}) implies the centrifugal
potential diminishes the role of trapping potential (in order to
stabilize the system, $\omega \leq \omega_0$). For simplicity, we
assume $\omega=\omega_0$ so the centrifugal force accurately
compensates the harmonic trapping potential in the $xy$ plane and
the system is uniform. The other part of rotation is included in
the first term whose role is identical to the Lorentz force acting
on a particle of charge $-e$ experiencing an effective magnetic
field $-B_\bot =\partial_x A_y -\partial_y A_x = -2m\omega/e$
along the $z$ axis and provides a structure of Landau level for
atoms. Hence at this point we can make an analogy between the
motion of charged particles with magnetic moment in a magnetic
field and neutral atoms also with magnetic moment in a rotating
frame and in an external magnetic field. In this respect, a
quantum gas of atoms confined in a harmonic trap rotating at the
critical frequency is analogous to an electron gas in a uniform
magnetic field except that the orbital and spin parts
independently couple to rotation and magnetic field for neutral
atom but not for the electrons. So there is one more degree of
freedom (rotation frequency) for neutral atom than electrons,
which makes our conclusion different from the de Haas-van Alphen
effect.

The Hamiltonian (\ref{1.2}) can be solved accurately as follows
\cite{ezawa}. Introducing the covariant momentum
\begin{eqnarray}
P_x=-i\hbar\partial_x+eA_x \, ,\,P_y=-i\hbar\partial_y+eA_y
\label{1.3}
\end{eqnarray}
and creation and destruction operators
\begin{eqnarray}
a=\frac{l_B}{\sqrt{2}\hbar}(P_x+iP_y) \,
,\,a^\dag=\frac{l_B}{\sqrt{2}\hbar}(P_x-iP_y) \label{1.4}
\end{eqnarray}
with $l_B=\sqrt{\hbar/eB_\bot}$ being magnetic length. Easily
proved $[a, a^\dag]=1$. So the Hamiltonian is diagonalized and the
energy spectrum is
\begin{eqnarray}
E_{p_z, n, \sigma_z}=\frac{p_z^2}{2m}+
(n+\frac{1}{2})\hbar\omega_c -\frac{g}{2}\mu_B B \sigma_z
\label{1.5}
\end{eqnarray}
with the cyclotron frequency being denoted by
$\omega_c=eB_\bot/m$.

After determining the energy spectrum of single particle, we
follow the method in \cite{abrikosov} to calculate the
thermodynamic potential
\begin{eqnarray}
\Omega=-\frac{eB_\bot}{(2\pi\hbar)^2 \beta}
\sum_{n=0}^{\infty}\sum_{\sigma_z} \int_{-\infty}^{\infty} dp_z
\ln[1+e^{-\beta(E_{p_z, n, \sigma_z}-\mu)}] \label{1.6}
\end{eqnarray}
with $\mu$ and $\beta$ representing chemical potential and inverse
temperature. By means of Poission sum formula we can find out the
oscillatory part of thermodynamic potential
\begin{eqnarray}
\Omega_{os}&=&-2Re\sum_{k=1}^{\infty}\sum_{\sigma_z}\Pi_{k\sigma_z}
\nonumber \\
\Pi_{k\sigma_z}&=&\frac{eB_\bot}{(2\pi\hbar)^2 \beta}
\int_0^\infty dx \int_{-\infty}^{\infty}dp_z e^{i2\pi kx}
\ln[1+e^{\beta(\mu_{\sigma_z}-E_{p_z, x)}}] \label{1.7}
\end{eqnarray}
with $\mu_{\sigma_z}=\mu+\frac{g}{2}\mu_B B \sigma_z$ and $E_{p_z,
x} =\frac{p_z^2}{2m}+ (x+\frac{1}{2})\hbar\omega_c$. $Re$
represents the real part. After completing integral over $x$,
$p_z$ and summation over $\sigma_z$
\begin{eqnarray}
\Omega_{os}=\frac{1}{2\pi^2
\beta}\left(\frac{m\omega_c}{\hbar}\right)^{3/2}\sum_{k=1}^{\infty}
\frac{1}{k^{3/2}\sinh{(2\pi^2k/\beta\hbar\omega_c)}}
\cos{\left(\frac{g\pi k\mu_B B}{\hbar\omega_c}\right)} \cos
{\left(\frac{2\pi k \mu}{\hbar\omega_c}-\frac{\pi}{4}\right)}
\label{1.8}
\end{eqnarray}
Before deciding the magnetic moment, it is necessary to determine
the chemical potential $\mu$ in terms of particle number $N$.
However as long as $\hbar\omega_c<<\mu$ and $g\mu_B B<<\mu$, the
relation between $N$ and $\mu$ is the same as for a free-electron
gas \cite{callaway}. According to above statement, we can consider
$\mu$ to be independent of the magnetic field. Therefore we attain
the magnetic moment $M_{os}=-\frac{\partial \Omega_{os}}{\partial
B}$
\begin{eqnarray}
M_{os}=\frac{eg}{4\pi \beta}
\left(\frac{m\omega_c}{\hbar^3}\right)^{1/2}
\sum_{k=1}^{\infty}\frac{1}{k^{1/2}\sinh{(2\pi^2k/\beta\hbar\omega_c)}}
\sin{\left(\frac{g\pi k\mu_B B}{\hbar\omega_c}\right)} \cos
{\left(\frac{2\pi k \mu}{\hbar\omega_c}-\frac{\pi}{4}\right)}
\label{1.9}
\end{eqnarray}
When we fix rotation frequency to equal with external trapping
frequency, the oscillation of magnetic moment is periodic in $B$
and the period is
\begin{eqnarray}
\Delta(B)=\frac{2\hbar\omega_c}{g\mu_B}=\frac{8m\omega}{eg}
\label{1.10}
\end{eqnarray}
and is proportional to rotation frequency. This is our main
conclusion. On the other hand, if $\omega_c=\frac{eB}{m}$ is
replaced into (\ref{1.8}), de Haas-van Alphen effect is recovered
and the magnetism oscillation is periodic in $B^{-1}$ with the
period $\Delta(B^{-1})=\frac{e\hbar}{m\mu}$. The different
oscillatory behavior apparently comes from contribution of
different cosine function in (\ref{1.8}). For de Haas-van Alphen
effect, the dependence on $B$ is cancelled accurately in the first
cosine function hence the oscillatory behavior comes from the
second cosine function. While for the atomic gases only the first
cosine function depends on the magnetic field $B$, so does
magnetic moment. From the physical point, different oscillatory
behavior consists in that the orbital motion of atom does not
couple to magnetic field and does not contribute to magnetic
moment so the whole magnetic moment only corresponds to the Pauli
paramagnetism.

Above only idealized atom gas is considered. But the influence of
atom-atom scattering is easily estimated \cite{abrikosov}. The
collisions between atoms give a finite lifetime for atomic states
$\tau$ which can be absorbed into energy spectrum with an extra
imaginary term $-i\hbar/2\tau$. One therefore expects that
collisions will lead to an additional exponential factor in
(\ref{1.8}) and (\ref{1.9}) $\exp{(-\pi k/\omega_c\tau)}$ which
suppresses the amplitude of the oscillation. At low temperature
since $\tau \sim T^{-2}$ \cite{smith} the exponential factor hence
the effect of collision can be neglected. When $\omega<\omega_0$,
the system is inhomogeneous and local density approximation can be
used. As a result, chemical potential $\mu$ in all formulae is
replaced by $\mu(x, y)
=\mu-\frac{1}{2}m(\omega_0^2-\omega^2)(x^2+y^2)$. Except near the
boundary $\mu(x, y)=0$, the conditions $\hbar\omega_c<<\mu(x, y)$
and $g\mu_B B<<\mu(x, y)$ are still satisfied specially in the
center of the trap potential. So our conclusion is still valid.

In conclusion, we calculate atomic Pauli paramagnetism in the
rotating two-component fermionic atomic gases and find that the
oscillatory behavior of magnetic moment is completely different
from de Haas-van Alphen effect as the function of magnetic field.
This distinction is due to that the orbital motion of atom does
not contribute to magnetic moment owing to atomic charge
neutrality.

Authors acknowledge support from NSFC Grant No.10675108.

\end{document}